\documentclass[final]{elsarticle}

\journal{Ultrasonics}

\usepackage{graphicx}
\usepackage{dcolumn}
\usepackage{bm}
\usepackage{siunitx}
\usepackage{amsmath,amsfonts,amssymb}
\usepackage{lineno,hyperref}
\usepackage[final]{changes}
\modulolinenumbers[1]

\colorlet{Changes@Color}{red}

\renewcommand{\vec}[1]{\bm{#1}}
\newcommand{\ee}{\mathrm{e}}
\newcommand{\ii}{\mathrm{i}}
\newcommand{\dd}{\mathrm{d}}

\begin{document}

\begin{frontmatter}
	
	\title{\added{Subwavelength beam focusing using a ball-shaped lens for ultrasound imaging}}
	
	\author[address1]{Victor H. S. Santos}%
	\author[address1]{Everton B. Lima}%
	\author[address1]{Andr\'e L. Baggio}
	\author[address2]{Jos\'e H. Lopes}%
	\author[address1]{Glauber T. Silva\corref{gtomaz@fis.ufal.br}}

	\address[address1]{Physical Acoustics Group,
		Instituto de F\'isica,
		Universidade Federal de Alagoas, 
		Macei\'o, AL 57072-970, Brazil}
	\address[address2]{Grupo de F\'isica da Mat\'eria Condensada, 
		N\'ucleo de Ci\^encias Exatas, Universidade Federal de Alagoas,
		Arapiraca, Alagoas 57309-005, Brazil}


\begin{abstract}
	\added{Ultrasonic superresolution images can be generated by means of (super) focusing acoustic beams to subwavelength dimensions or using algorithm-based methods. Here, we demonstrate that ultrasonic pulses which are superfocused by a ball-shaped lens can be used to produce	superresolution images.}
	The imaging system is comprised of a circular flat transducer with an operation frequency of \added{$\SI{1}{\mega\hertz}$}, and a ball lens centered in the beam axis of symmetry.
	The corresponding wavelength in water is $\lambda_0=\SI{1.53}{\milli\meter}$.
	The system resolution is $0.6\lambda_0$ in the focal plane at one wavelength away from the lens.
	The superresolution method is compared with a conventional ultrasonic system \added{based on a spherically focused transducer.}
	Our method presents twice more resolution with a shorter depth-of-field of $2\lambda_0$.
	\added{Possible applications that take advantage of these features are discussed, as well as some limitations of the proposed technique.}
\end{abstract}


\end{frontmatter}



\section{Introduction}
The spatial resolution of ultrasonic imaging systems is primarily restricted by the diffraction limit (i.e., beam focusing in a disk roughly with one wavelength diameter)~\cite{Kino1987}.
This effect thwarts subwavelength focusing at a given frequency. 
A finer resolution can  be achieved by increasing the frequency at the expense of less ultrasound penetration due to absorption and also a high-cost electronics.

Different methods have overcome the ultrasonic diffraction limit.
Notably,  harmonic generation was used to make images at a doubled frequency, which means an improvement of 100\% is the spatial resolution~\cite{Rugar1984,Ward1997}.
Nonlinear ultrasonic wave mixing forms images at the difference-frequency with the fundamental high-frequency resolution~\cite{Fatemi1998, Silva2006, Baggio2017}.
\added{Also, the nonlinear wave interaction gives rise to a sum-frequency component (i.e., the sum of the fundamental frequencies)  that can be used to form superresolution images~\cite{Mitri2007}.}
Another approach uses the time-reversal wave phenomenon to focus an ultrasound beam~\cite{Fink2000}.
Lenses made of a phononic crystal~\cite{Sukhovich2009} and metamaterial~\cite{Zhu2011} also showed promising results on focusing ultrasonic beams beyond the diffraction limit.
By using ultra-fast aquisitions based on plane wave transmissions at the rate of a thousand frames per second, the image of ultrasound contrast agents (microbubbles) in echography surpassed the diffraction limit by more than a tenfold~\cite{Errico2015}.
\added{Algorithm-based methods that aim at reducing the dependence of resolution on pulse shape and width can also produce superresolution images~\cite{Fan2014,George2019}.}
Despite early success, these approaches may involve complex material engineering, low efficiency, or intense signal processing algorithms.
Desirably,
superresolution methods should have scalability varying with the wavelength with relatively simple electronics to find practical applications in biomedical imaging, nondestructive testing, and acoustic microscopy.

Recently, it has been experimentally demonstrated that a polymer ball-shaped lens can focus an ultrasonic beam beyond the diffraction limit~\cite{Lopes2017}.
The beam width can be even smaller than half-wavelength with a depth-of-field of few wavelengths.
The focal region takes place in the lens shadow region, \added{which is centered at one wavelength (or more) away from the lens}, and the resulting wave is non-evanescent.  
Some preliminary numerical results on ball lens superfocusing have been investigated in Ref.~\cite{Minin2017}.
Superresolution was also achieved by employing a cylindrical-shaped lens~\cite{Canle2019}.
These features are particularly suitable for enhancing the imaging performance of ultrasonic systems.

We present here a superresolution ultrasonic (SU) imaging system based on the ball lens focusing method.
The system is designed to operate in the pulse-echo mode and to form C-scan images.
We show that the SU system has a superior resolution and depth-of-field compared to an ultrasonic system, \added{which utilizes a spherically focused transducer.}

\section{Methods and materials}

\subsection{Superresolution image formation}
\begin{figure}
	\begin{center}
		\includegraphics[scale=.9]{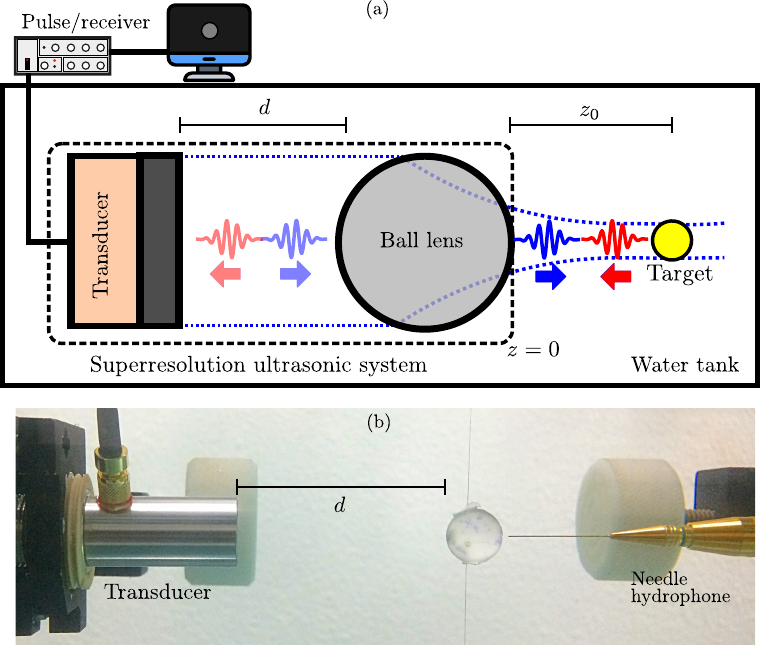}
	\end{center}
	\caption{ (a) Schematics of the superresolution ultrasonic (SU) imaging system.
		The incident (blue) pulse hits a target object at the focal plane $z=z_0$.
		The system receives the backscattered pressure (red pulse).
		The acquired signal is post-processed and displayed on a computer.
		(b) Photography of the SU system. 
	}
	\label{fig:system}
\end{figure}

Consider ultrasonic waves propagating in a homogeneous non-viscous fluid of density $\rho_0$ and speed of sound $c_0$.
The acoustic pressure is described at position $\bm r$, concerning the coordinate system set in the right-hand pole of the ball lens (see Fig.~\ref{fig:system}), and time $t$.

We assume that the active surface of the transducer vibrates uniformly with normal velocity denoted by $v_n$.
For a linear and time  invariant system, the transmitted pressure reads\cite{Stepanishen1981}
\begin{equation}
\label{ptr}
p_\text{tr}(\vec{r},t) = \rho_0 v_n(t) * \partial_t h(\vec{r},t),
\end{equation}
where $\partial_t$ is time derivative, $h$ is the spatial impulse response of the transducer and asterisk  means convolution in time--see Eq.~\eqref{timeconvolution}.
The vibration velocity is expressed by
\begin{equation}
\label{vn}
v_n(t) = v_\text{exc}(t) * e_\text{tr} (t),
\end{equation}
where $v_\text{exc}(t)$ is the excitation voltage and $e_\text{tr} (t)$ is the electromechanical impulse response of the transducer in the transmitting mode.

The lens focuses the transmitted pressure through a scattering process~\cite{Lopes2017}.
In turn, ultrasound scattering can be model as a linear and shift-invariant system.
The incident pressure $p_\text{in}$ to a target object can be expressed as the spatial convolution--see Eq.~\eqref{spaceconvolution}--between the transmitted pressure by the transducer and the lens spatial response function $h_\text{lens}$,
\begin{equation}
\label{pin}
p_\text{in}(\vec{r},t) =   p_\text{tr} ({\bm r},t) \star h_\text{lens}(\vec{r}).
\end{equation}
The backscattered pressure  by the target is given by~\cite{Jensen1991}
\begin{equation}
p_\text{sc}(\vec{r},t) =   p_\text{in} ({\bm r},t) \star f(\vec{r}), 
\end{equation}
with $f$ being the object's function.
The scattered wave is also focused by the  lens yielding the pressure to be received  by the transducer,
\begin{equation}
\label{prec}
p_\text{rec}(\vec{r},t) =   p_\text{sc} ({\bm r},t) \star h_\text{lens}(\vec{r}).
\end{equation}
The  signal in the transducer terminals is described by~\cite{Stepanishen1981}
\begin{equation}
s(t) =   e_\text{rec}(t)* p_\text{rec} ({\bm r},t)  * h(\vec{r},t),
\end{equation}
\added{where $e_\text{rec} (t)$ is the electromechanical impulse response of the transducer in the receiving mode.}
Substituting Eqs.~\eqref{vn}--\eqref{prec} into this equation yields
\begin{subequations}
	\begin{align}
	\label{vrec}
	s(t) &=  v_\text{pe}(t) *  \partial_t  g(\vec{r},t)
	* g(\vec{r},t)\star f(\vec{r}),\\
	g(\vec{r},t) &= h(\vec{r},t)\star h_\text{lens}(\vec{r}),\\
	v_\text{pe}(t) &= \rho_0 v_\text{exc}(t) * e_\text{tr} (t)*e_\text{rec}(t).
	\end{align}
\end{subequations}
Equation~\eqref{vrec} has a similar structure of Eq.~(45) in Ref.~\cite{Jensen1991}, which describes a conventional ultrasonic pulse-echo system.
The spatial impulse function of the SU system $g(\bm r,t)$ accounts for  diffraction effects of the transducer and lens.
The pulse-echo wavelet $v_\text{pe}(t)$  includes the voltage excitation and the transducer electromechanical impulse response.

According to Eqs.~\eqref{fourier}, \eqref{dfourier} and \eqref{fconvolution}, the detected signal can be expressed in the frequency-domain as
\begin{equation}
\label{Sw}
S({\bm r},\omega) =\ii \rho_0 \omega  V_\text{exc}(\omega)E_\text{tr}(\omega) E_\text{rec}(\omega)G^2({\bm r},\omega)\star f(\vec{r}).
\end{equation}
Hereafter, an uppercase function denotes the Fourier transform of its corresponding time-domain counterpart.
From Eq.~\eqref{pin}, we see that the incident pressure in the frequency-domain is
\begin{equation}
\label{Pin}
P_\text{in}({\bm r},\omega)= -\ii \omega \rho_0 V_\text{exc}(\omega)E_\text{tr}(\omega) G({\bm r}, \omega).
\end{equation}
Combining Eqs.~\eqref{Sw} and \eqref{Pin} results
\begin{equation}
\label{SS}
S({\bm r},\omega) = A(\omega)P_\text{in}^2({\bm r},\omega)\star f(\vec{r}),
\end{equation}
where $A(\omega) = [\ii \rho_0 \omega  V_\text{exc}(\omega)E_\text{tr}(\omega)]^{-1}.$

The SU imaging system is considered as a spatially invariant system.
The image formation is then described by the point spread function (PSF).
This function describes how what should be a point target is spread out by diffraction.
We determine the PSF by assuming that the object function is a Dirac delta distribution 
\begin{equation}
f(\vec{r})=\delta(\vec{r}'-\vec{r}).
\end{equation}
Substituting this function into Eq.~\eqref{SS} and following Eq.~\eqref{gdelta}, we find
the detected signal of a point target as
\begin{equation}
\label{SS2}
S_\delta(\vec{r},\omega) = A(\omega)P_\text{in}^2(\vec{r},\omega).
\end{equation}
The SU PSF can be defined as the detected signal at the focal plane ($z=z_0$) divided by the same signal in the focus point
both at the center-frequency $\omega=\omega_0$, 
\begin{align}
\label{hPSF}
h_\text{PSF}(x,y)&\equiv  \left|\frac{S_\delta(x,y,z_0,\omega_0)}{S_\delta(0,0,z_0,\omega_0)}\right|=
\left|\frac{P_\text{in}(x,y,z_0,\omega_0)}{P_\text{in}(0,0,z_0,\omega_0)}\right|^2.
\end{align}
Here we have also used Eq.~\eqref{SS2}.
Finally, the image of an object is given by 
\begin{equation}
i(x,y) = h_\text{PSF}(x,y)\star f(x,y) + n(x,y),
\end{equation}
where $f(x,y)$ is the 2D object function and 
$n(x,y)$ is the system noise. 
\added{Albeit after determining the PSF the SU images can be further enhanced by deconvolution algorithms~\cite{Perciano2013}, the proposed method is not an algorithm-based technique~\cite{Zhao2016}.}

\added{We shall compare the superresolution method with conventional ultrasonic (CU) technique that employs a spherically focused transducer.}
The spatial resolution of the CU system is defined as the beam intensity at full width at half maximum (FWHM)~\cite{Kino1987},
\begin{equation} 
\label{wcu}
w_\text{CU} = 1.02 \lambda_0 N =1.13 \lambda_0,
\end{equation}
where $N$ is the transducer f-number.
The CU depth-of-field is described as the full depth at half maximum (FDHM) of the axial beam intensity~\cite{Kino1987},
\begin{equation}
\label{dcu}
d_\text{CU} =7.1 \lambda_0 N^2 = 8.74 \lambda_0. 
\end{equation}
\begin{figure*}
	\begin{center}
		\includegraphics[scale=.85]{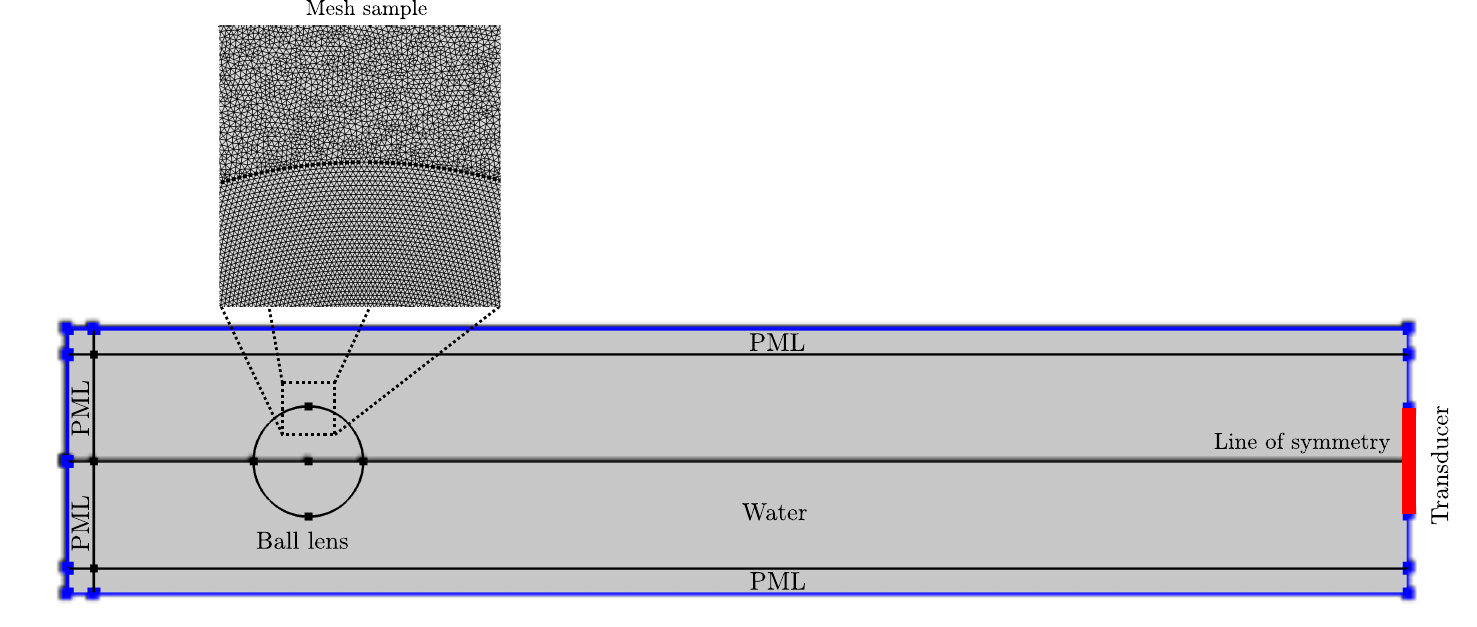}
	\end{center}
	\caption{\label{fig:mesh} 
		Description of the domains used in the finite element simulation of the ultrasonic pulse superfocusing.}
\end{figure*}

\subsection{Finite element simulations}
We now explain how to obtain the system PSF with numerical simulations.
The incident pressure $p_\text{in}$ is computed in time-domain using the finite element method in Comsol Software (Comsol Inc., USA).
We also assume that the transducer is a circular rigid piston with the normal vibration velocity given by a Gaussian vibration packet,
\begin{equation}
v_n (t)= v_0 \ee^{-(t-t_\text{c})^2/\Delta t^2} \sin \omega_0 t,
\end{equation}
where $v_0$ is the oscillation amplitude, $\Delta t$ and $t_\text{c}$ are the pulse width and  time delay, respectively. 
The pressure $P_\text{in}$  is calculated via the fast Fourier transform algorithm.
The PSF is then computed through Eq.~\eqref{hPSF}.
The simulation domain and mesh are described in Fig.~\ref{fig:mesh}. 
The propagation medium is water, which is assumed to be an inviscid fluid.
The ball lens is made of rexolite polymer.
In the simulations, we used the Transient Pressure Acoustics and Solid Mechanics modules.
The numerical boundary conditions are the Perfect Match Layer (PML) combined with the plane wave radiation condition.
The simulation parameters are summarized in Table \ref{tab:parameters}.
\begin{table}
	\caption{\label{tab:parameters}
		Parameters of the numerical simulation at room temperature and pressure, and computational information.
	}
	\begin{tabular}{lr}
		\hline
		\hline
		\textbf{Parameter} & \textbf{Value}\\
		\hline
		\textbf{Medium (water)} &\\
		Density ($\rho_0$) &  $ \SI{1000}{\kilogram \per \meter\cubed}$\\
		Speed of sound ($c_0$) & $ \SI{1480}{\meter \per \second}$\\
		Wavelength ($\lambda_0$) & $\SI{1.53}{\milli\meter}$\\
		Mesh element size $\lambda_0/11$ & $\SI{89.6}{\micro\meter}$\\
		Dimensions (free triangular mesh) & $\SI{12.24}{\milli\meter}$ (W) $\times$ $\SI{150.8}{\milli\meter}$ (L)\\ 
		PML (mapped mesh) & $\SI{3.06}{\milli\meter}$ (W) $\times$ $\SI{3.06}{\milli\meter}$ (L)\\
		\hline
		\textbf{Transducer} \\ 
		Diameter  & $\SI{12}{\milli\meter}$\\
		Pressure at the active surface ($p_0$) & $\SI{1.5}{\mega\pascal}$\\
		Normal velocity ($v_0$) & $\SI{1}{\meter\per\second}$\\
		Pulse width ($\Delta t$) & $\SI{2.06}{\micro\second}$\\
		Time delay ($t_0$) & $\SI{6.18}{\micro\second}$\\
		Center frequency ($f_0$) & $\SI{1}{\mega\hertz}$\\
		Sampling frequency ($f_\text{s}$) & $\SI{5.25}{\mega\hertz}$\\
		\hline
		\textbf{Ball lens (rexolite 1422)}& \\
		Diameter   &    $\SI{12}{\milli\meter}$\\
		Distance to the transducer ($d$) & $\SI{120}{\milli\meter}$\\
		Density  &  $ \SI{1049}{\kilogram \per \meter\cubed}$\\
		Longitudinal speed of sound  & $ \SI{2337}{\meter \per \second}$\\
		Shear speed of sound  & $ \SI{1157}{\meter \per \second}$\\
		Mesh element size & $\SI{70.1}{\micro\meter}$\\
		\hline
		{\bf Additional information }& \\
		Computation time & \SI{53}{\hour}\\
		CPU  &      E5-2690 3.00GHz, 20 cores\\
		Operating system &        Linux\\
		\hline
	\end{tabular}
\end{table}

We have performed the mesh convergence analysis for the simulated superresolution system.
The two parameters in this analysis are the full width at half maximum (FWHM) and full depth at half maximum (FDHM).
We computed the relative error of these parameters by varying the mesh element size $\lambda_0/n$, for which  $n=5,7,9,11$.
The relative error is defined as
\begin{equation}
\epsilon = \left|1 - \frac{x_n}{x_{11}}\right|,
\end{equation}
where $x_n=\text{FWHM}, \text{FDHM}$ with the corresponding number of points per sampling wavelength.
The correct value in the error computation is assumed to be $x_{11}$.
For $n=9$ the errors are below $0.5\,\%$.
\begin{figure}
	\begin{center}
		\includegraphics[scale=.5]{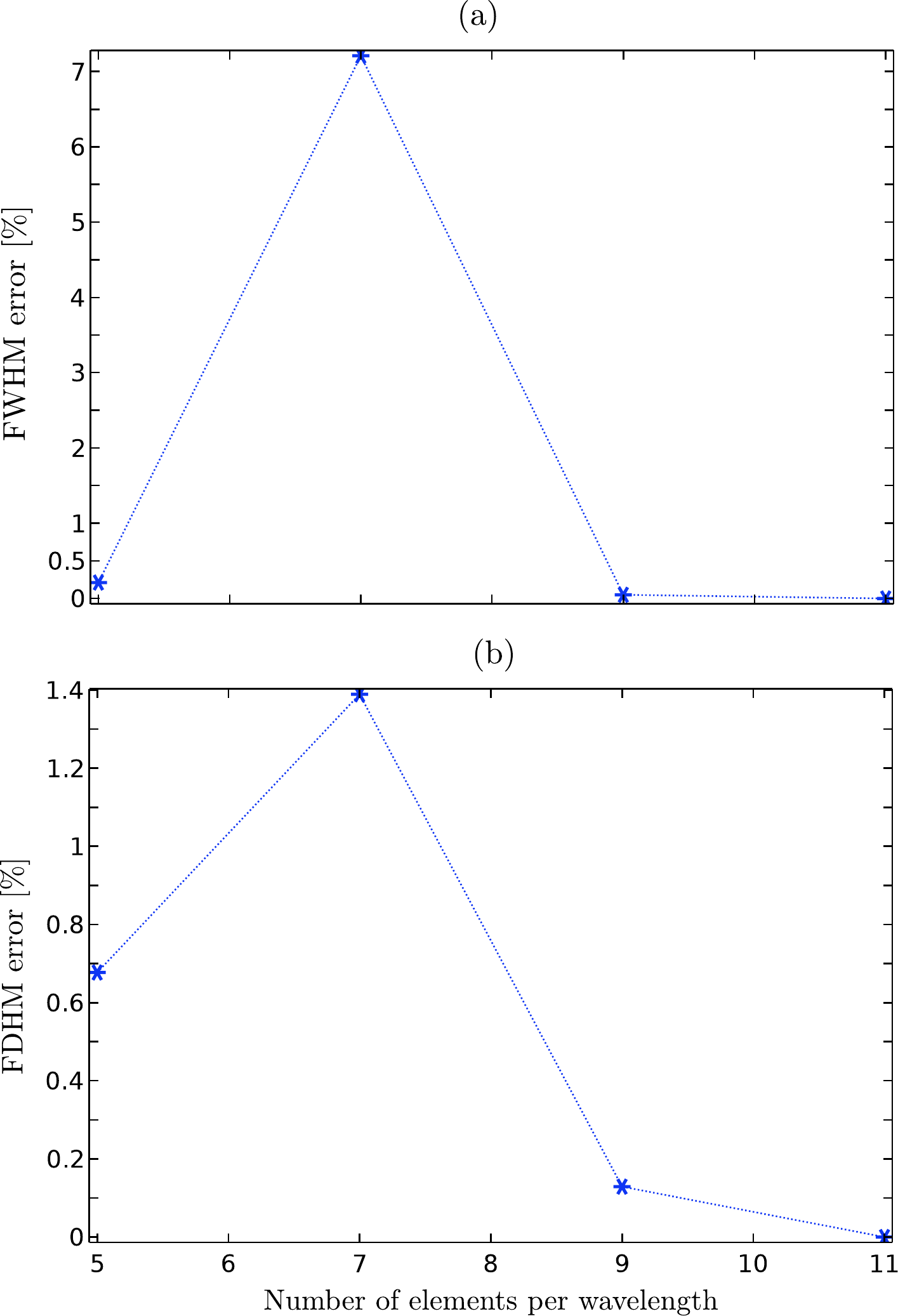}
	\end{center}
	\caption{\label{fig:error} 
		Error analysis of the mesh convergence (a) FWHM and (b) FDHM.
	}
\end{figure}

\subsection{Superresolution ultrasonic system}
The SU imaging system is depicted in Fig.~\ref{fig:system}.
It is composed of a  circular flat transducer (ISG014SM, NdtXducer LCC, USA), with a diameter of $\SI{12}{\milli\meter}$ and center frequency of $\omega_0/2\pi = \SI{1}{\mega\hertz}$.
A ball-shaped lens made of rexolite with a $\SI{12.2}{\milli\meter}$-diameter is suspended at $d=\SI{120}{\milli\meter}$ away from the transducer active element along its central axis. 
\added{At this distance, the incident beam intensity from the transducer reaches its maximum.}
\added{Rexolite material was chosen due to its low attenuation and  acoustic impedance close to water.}
These parameters are chosen to attain a spatial resolution close to half-wavelength and a depth-of-field of few wavelengths~\cite{Lopes2017}.
The experimental apparatus is immersed in a water tank with dimensions of $\SI{109}{\centi\meter}$ (L) $\times$ $\SI{54}{\centi\meter}$ (W) $\times$ $\SI{57}{\centi\meter}$ (H).
The characteristic wavelength of a pulse is $\lambda_0=\SI{1.53}{\milli\meter}$.
The transducer is driven by a pulse/receiver (DPR300, JSR Ultrasonics, USA) with negative spike pulses (excitation voltage of $\SI{200}{\volt}$, damping of $\SI{333}{\ohm}$, the energy of $\SI{16}{\milli\joule}$, and repetition frequency of $\SI{100}{\hertz}$).
The emitted pulse is focused in the shadow region of the lens at $z=z_0=1.04\lambda_0=\SI{1.6}{\milli\meter}$.
This defines the imaging plane of the system.
The peak pressure in the system focus of $\SI{7.6}{\kilo\pascal}$ was measured by a $\SI{0.2}{\milli\meter}$ needle hydrophone (NH0200, Precision Acoustics, UK).
The backscattered pressure by a target object is re-focused by the ball lens and acquired by the transducer.
The corresponding signal is pre-processed with a low-pass filter  ($\SI{3}{\mega\hertz}$-cutoff frequency and $\SI{43}{\decibel}$-gain), digitized by a 12-bit A/D converter with $\SI{60}{\mega Sample\per\second}$ (PCI-5105, National Instruments, USA), and gated in a $\SI{4}{\micro\second}$-time window. 
The image pixel corresponds to the magnitude of the acquired signal in the frequency-domain at $\SI{1}{\mega\hertz}$.
The SU beam raster-scans the object in steps of $\SI{150}{\micro\meter}$ that is much smaller than the wavelength.

\begin{figure}
	\begin{center}
		\includegraphics[scale=.6]{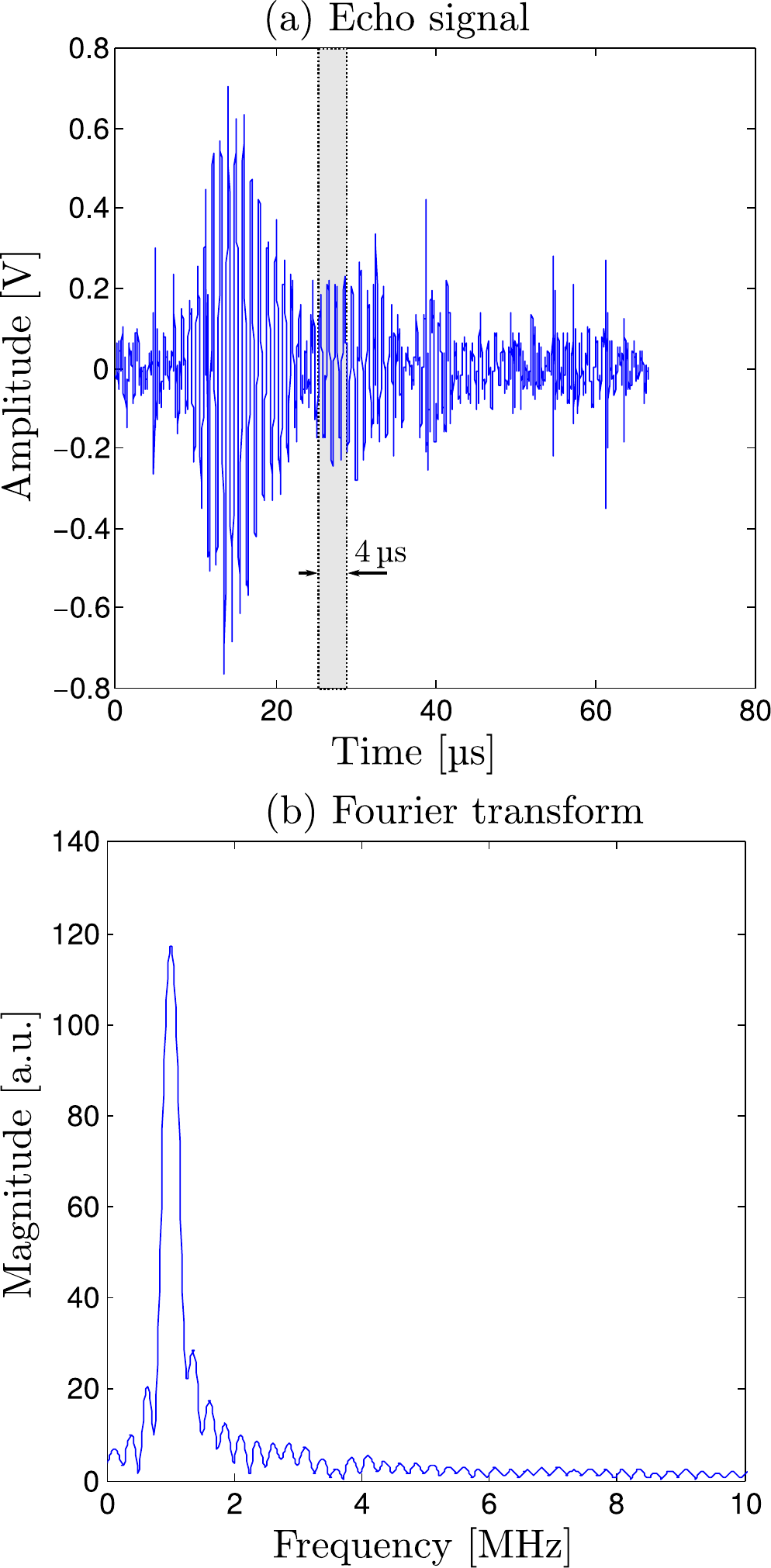}
	\end{center}
	\caption{\label{fig:echo} 
		(a) Detected echo from a point target placed at the system focus $(0,0,\SI{3}{\milli\meter})$.
		(b) Fourier transform of the echo signal gated in a $\SI{4}{\micro\second}$-interval depicted by the gray region. 
	}
\end{figure}

A typical echo signal measured as the response to a point target (e.g., the tip of a needle with a $\SI{125}{\micro\meter}$-diameter)
placed at the system focus $(0,0,\SI{1.6}{\milli\meter})$ is shown in Fig.~\ref{fig:echo}, panel (a).
The Fourier transform of the gated echo at a $\SI{4}{\micro\second}$-window is depicted in panel (b). 

\section{Results and discussion}
\begin{figure*}
	\begin{center}
		\includegraphics[scale=.65]{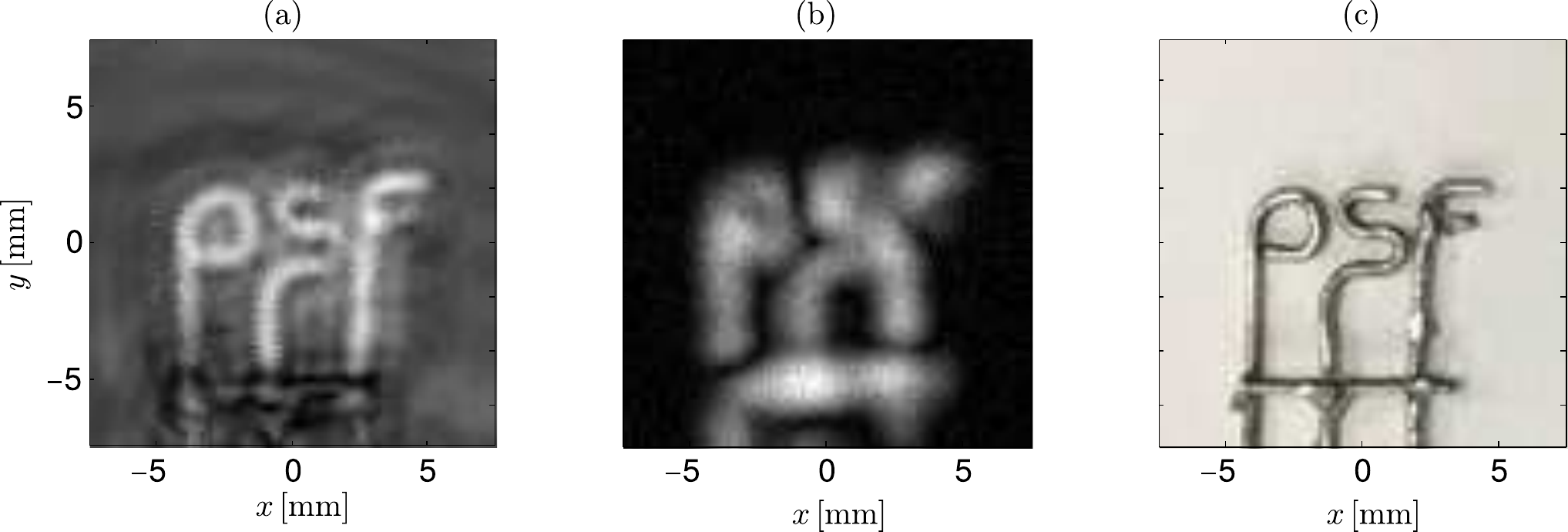}
	\end{center}
	\caption{\label{fig:imagePSF}
		Images of the letters `PSF' made of thin wires with
		a $\SI{0.5}{\milli\meter}$-diameter. (a) Superresolution ultrasonic image \added{with a dynamic range of $\SI{30}{\decibel}$.} 
		(b) Conventional ultrasonic image \added{by a spherically focused transducer with a dynamic range of $\SI{16}{\decibel}$.}
		\added{Both images were obtained at  $\SI{1}{\mega\hertz}$.}
		(c) Photography.}
\end{figure*}

In Fig.~\ref{fig:imagePSF}, we show images of letters `PSF' made of thin wires with
a $\SI{0.5}{\milli\meter}$-diameter: panel (a), (b), and (c) show, respectively, the SU, conventional ultrasonic (CU), and photography. 
The CU system is based on a spherically {focused} transducer with a diameter of $\SI{45}{\milli\meter}$, $\SI{50}{\milli\meter}$-focal distance (image plane), and f-number $N= 1.11$.
The transducer operates in the pulse-echo mode at $\SI{1}{\mega\hertz}$.
The same electronic hardware and signal processing are used for both the SU and CU systems.
The SU and CU images have $100\times100$ pixels.
These images differ in several important ways.
The letters can be seen in the SU image, while 
they are not recognized in the CU image.
The horizontal bar is shown in panel (c) is placed $\SI{0.5}{\milli\meter}$ in front of the vertical wires. 
The bar is visible in the CU image, whereas it is not seen in the SU image.
\added{Moreover, the SU and CU images have dynamic range of $\SI{30}{\decibel}$ and $\SI{16}{\decibel}$, respectively.}

\begin{figure}
	\begin{center}
		\includegraphics[scale=.4]{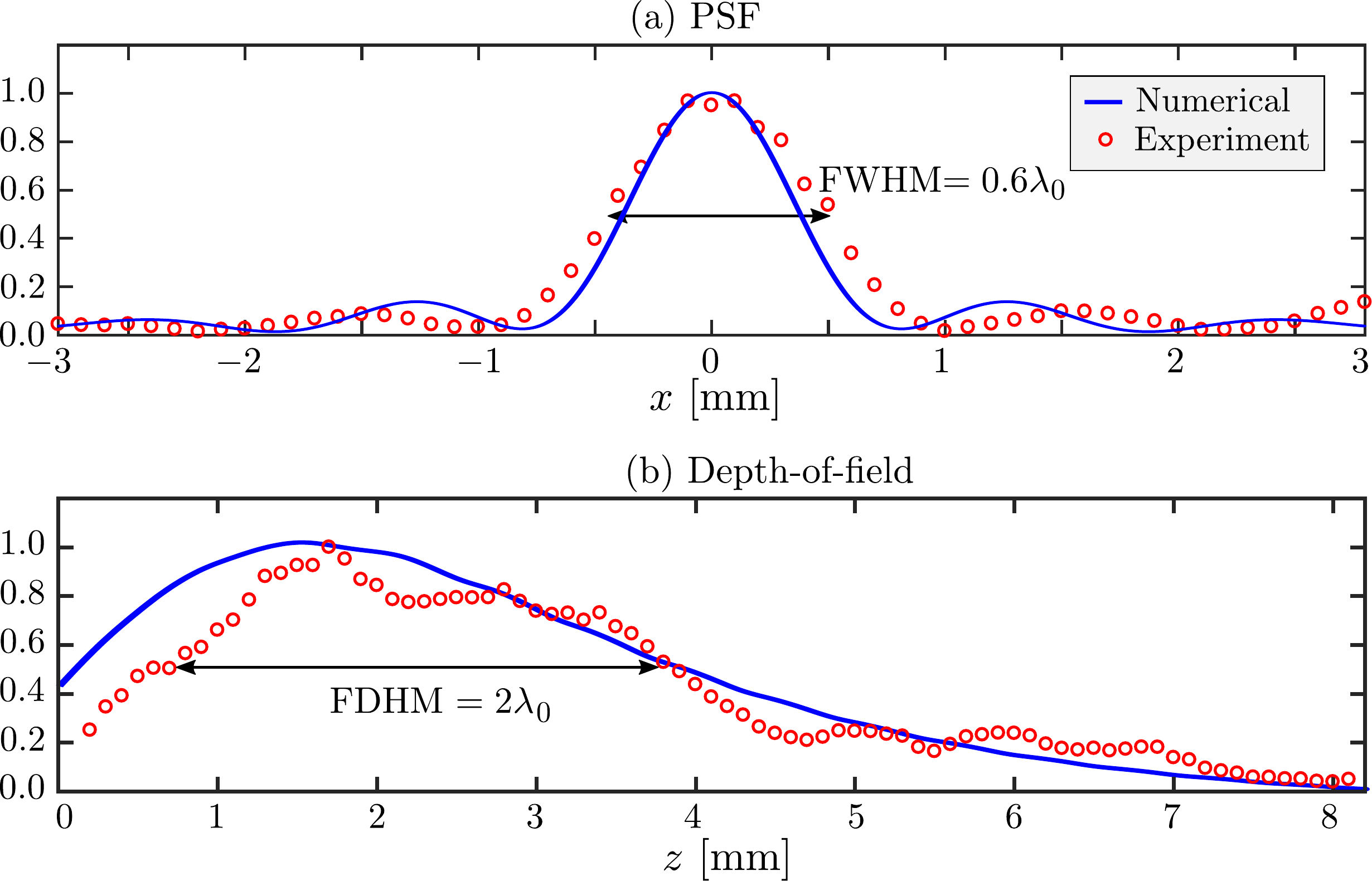}
	\end{center}
	\caption{\label{fig:1DPSF} 
		Comparison between the experimental and numerical results. 
		(a) SU PSF.
		(b) Normalized SU depth-of-field.      
		The SU system operates in the pulse-echo mode with $\SI{1}{\mega\hertz}$-center frequency. 
	}
\end{figure}

In Fig.~\ref{fig:1DPSF}, we compare the numerical simulation and experimental results.
The physical parameters used in the numerical simulation are summarized in Table~\ref{tab:parameters}.
The target object is the tip of a needle with a diameter of $\SI{125}{\micro\meter}$,
which is much smaller than the characteristic wavelength $\lambda_0=\SI{1.53}{\milli\meter}$.
Panel (a) shows the SU PSF.
The FWHM of the experimental and numerical PSF are $0.6\lambda_0 = \SI{0.91}{\milli\meter}$
and $0.5\lambda_0 = \SI{0.76}{\milli\meter}$, respectively.
From Eq.~\eqref{wcu}, we see that the FWHM of the CU system is $1.13 \lambda_0=\SI{1.70}{\milli\meter}$.
Panel (b) illustrates the depth-of-field of the SU system.
We see then that the experimental and numerical FDHM are $2\lambda_0 = \SI{3.06}{\milli\meter}$
and $2.13\lambda_0 = \SI{3.27}{\milli\meter}$, respectively.
Referring to Eq.~\eqref{dcu}, the experimental FDHM is four times smaller than the CU FDHM.
We have also numerically calculated the focusing gain of the lens (i.e., the ratio in dB of the pressure at focus to transmitted pressure at $z=-\SI{12}{\milli\meter}$) to be $\SI{13}{\decibel}$.
For comparison, the theoretical gain of the CU system is $\SI{26}{\decibel}$~\cite{Kino1987}.

\begin{figure}[t]
	\begin{center}
		\includegraphics[scale=1]{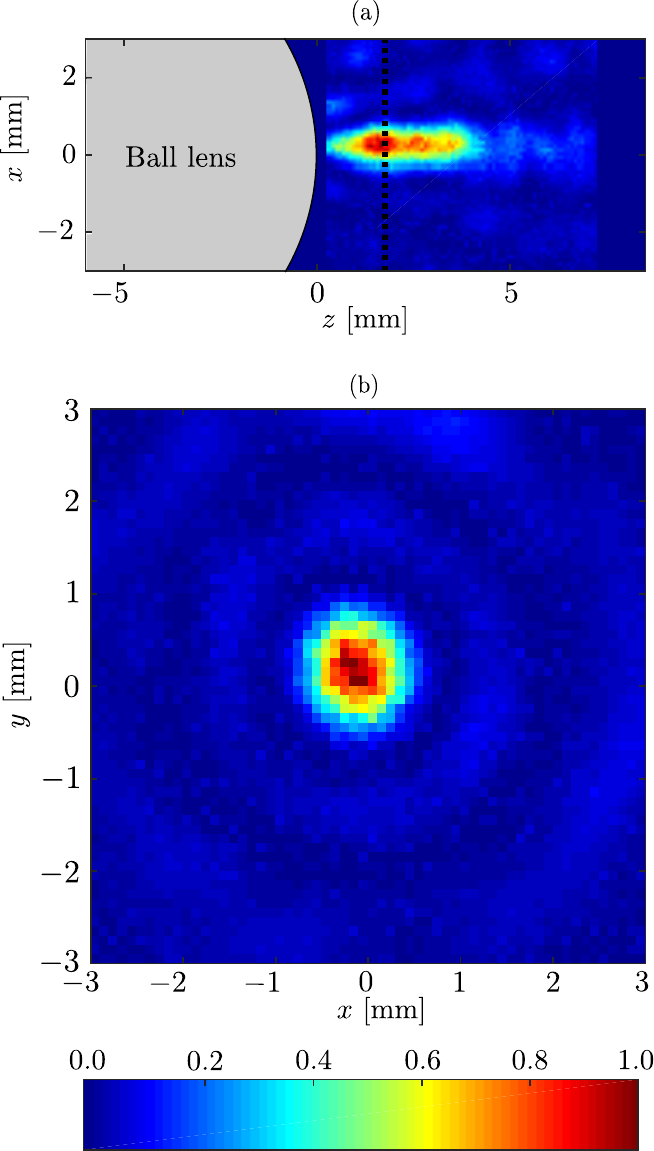}
	\end{center}
	\caption{(a) Normalized depth-of-field of the SU system.
		The vertical dotted line in (a) indicates the focal plane at $z=\SI{1.6}{\milli\meter}$.
		(b) System PSF. 
		\label{fig:PSF2D}}
\end{figure}

In Fig.~\ref{fig:PSF2D}, panel (a) displays the experimental depth-of-field image of the SU system.
The PSF image is presented in panel (b).
Moreover, the ultrasonic pulse propagation is illustrated by the video in the Supplementary material.
The propagating pressure and axial component stress tensor are shown outside and inside the lens, respectively.
It is worth noticing the pulse build-up inside the lens due to the focusing effect.
Subsequently, the pulse is transmitted into the surrounding liquid.
\added{A rigid sphere of diameter $\lambda_0$ is placed at the system focus. 
	The sphere scatters the incident pulse, and the echo propagates backwardly towards the transducer.} 

\added{In respect to the features mentioned above, superfocused beams can be used in ultrasound biomicroscope (UBM) that is employed in ophthalmic imaging~\cite{Silverman2009}.
	While this technology requires a frequency above $\SI{35}{\mega\hertz}$, the superfocusing method would need half of that frequency to produce images with similar lateral resolution.
	For instance, consider a superresolution system with frequency of $\SI{25}{\mega\hertz}$,
	wavelength of $\lambda_0=\SI{60}{\micro\meter}$, and lens diameter of $20 \lambda_0 = \SI{1.2}{\milli\meter}$.
	According to~\cite{Lopes2017}, this systen achieves a lateral resolution of $0.6\lambda_0 = \SI{36}{\micro\meter}$, depth-of-field of $6.5\lambda_0 = \SI{390}{\micro\meter}$,
	and focal distance of $7\lambda_0 = \SI{420}{\micro\meter}$.
	These parameters are compatible with those of a UBM with a $\SI{50}{\mega\hertz}$-center frequency~\cite{Silverman2009}.
	The obtained depth-of-field makes the superfocused system more suitable for C-scan images.}

\added{The main limitations of the SU method using a ball lens is having a relatively low contrast (dynamic range of $\SI{30}{\decibel}$) and a fixed focus. 
	Post-processing amplification and improving focus gain by the lens may increase the dynamic range.
	The beam steering of the SU system has to be done mechanically.
	Combining electronic beam steering with the proposed method is yet to be developed.}

\section{Summary and conclusions}
In summary, we have introduced a superresolution ultrasonic (SU) imaging system composed of a circular flat
transducer and a ball-shaped lens.
Excellent agreement is found between numerical simulations of the system PSF and experimental data.
The SU system has nearly a half-wavelength ($0.6 \lambda_0$) spatial resolution.
The images are formed in the nearfield at $1.04\lambda_0$ away from the lens with a depth-of-field of $2\lambda_0$.
These features add substantial improvements to SU images compared with conventional ultrasonic systems, i.e., higher resolution and sharper focusing.

The SU method may find fruitful applications for diagnostic imaging of skin and eye, which requires shallow image scanning, and detecting near-surface flaws in materials. 
Also, this method might be suitable for enhancing acoustic microscopy resolution while keeping the same operational frequency and
electronic hardware.

\section*{Acknowledgments}
{G. T. Silva thanks the National Council for Scientific and Technological
Development--CNPq, Brazil (Grant No. 307221/2016-4) for financial support.}

\appendix
\section{Mathematical background}
We summarize here some mathematical expressions used in the main text.
For a time-dependent function denoted by $g(t)$,
the Fourier transform  is given by
\begin{equation}
G(\omega)= \mathcal{F}[g(t)] = \int_{-\infty}^{\infty} g(t)
\ee^{-\ii \omega t}\, \dd t,
\end{equation} 
where $\omega$ is angular frequency.
The inverse Fourier transform reads
\begin{equation}
\label{fourier}
g(t)= \mathcal{F}^{-1}[G(\omega)]=\frac{1}{2\pi} \int_{-\infty}^\infty 
G(\omega) \ee^{\ii \omega t}\, \dd \omega.
\end{equation}
The Fourier transform of a time derivative of a function is given by
\begin{equation}
\label{dfourier}
\mathcal{F}\left[\frac{\dd g}{\dd t}\right] = \ii \omega \mathcal{F}\left[g(t)\right]=\ii \omega G(\omega).
\end{equation}
The convolution in time between two functions $g_1$ and $g_2$ is defined as
\begin{equation}
\label{timeconvolution}
g_1(t) * g_2(t) = \int_{-\infty}^\infty g_1(t')g_2(t-t') \, \dd t'.
\end{equation}
While the spatial convolution between two functions of configuration space $g_1(\vec{r})$ and $g_2(\vec{r})$ is expressed by
\begin{equation}
\label{spaceconvolution}
g_1(\vec{r}) \star g_2(\vec{r}) = 
\int_{\mathbb{R}^3} g_1(\vec{r}) g_2(\vec{r}-\vec{r}')\,\dd V',
\end{equation}
where $\dd V'$ is the volume element.
In the $xy$-plane, the convolution is reduced to
\begin{equation}
g_1(x,y) \star g_2(x,y) = 
\int_{-\infty}^\infty \int_{-\infty}^\infty g_1(x,y) g_2(x-x',y-y')\,\dd x' \, \dd y',
\end{equation}
The convolution between a function $g(\vec{r})$ and the Dirac delta distribution $\delta(\vec{r})$
results
\begin{equation}
\label{gdelta}
g(\vec{r}) \star \delta(\vec{r}) = g(\vec{r}).
\end{equation}
The Fourier transform of a convolution in time of two functions is given by
\begin{equation}
\label{fconvolution}
\mathcal{F}\left[g_1(t)*g_2(t)\right] = \mathcal{F}\left[g_1(t)\right]
\mathcal{F}\left[g_2(t)\right] =
G_1(\omega)G_2(\omega).
\end{equation}

\section*{References}	

\end{document}